\documentclass[aps,prl,preprint,superscriptaddress]{revtex4}
\usepackage{graphicx}
\usepackage{color}

\begin{document}

\title{Extracting the Temperature of Hot Carriers in Time- and Angle-Resolved Photoemission}
\author{S\o ren Ulstrup$^{\ast}$}
\affiliation{Department of Physics and Astronomy, Interdisciplinary Nanoscience Center (iNANO), Aarhus University, Denmark}
\author{Jens Christian Johannsen$^{\ast}$}
\affiliation{Institute of Condensed Matter Physics, \'Ecole Polytechnique F\'ed\'erale de Lausanne (EPFL), Switzerland}
\author{Marco Grioni} 
\affiliation{Institute of Condensed Matter Physics, \'Ecole Polytechnique F\'ed\'erale de Lausanne (EPFL), Switzerland}
\author{Philip Hofmann}
\affiliation{Department of Physics and Astronomy, Interdisciplinary Nanoscience Center (iNANO), Aarhus University, Denmark}
\affiliation{email: philip@phys.au.dk \\ $^{\ast}$ These authors contributed equally to the work. }

\date{\today}
 \begin{abstract}

The interaction of light with a material's electronic system creates an out-of-equilibrium (non-thermal) distribution of optically excited electrons. Non-equilibrium dynamics relaxes this distribution on an ultrafast timescale to a hot Fermi-Dirac distribution with a well-defined temperature. The advent of time- and angle-resolved photoemission spectroscopy (TR-ARPES) experiments has made it possible to track the decay of the temperature of the excited hot electrons in selected states in the Brillouin zone, and to reveal their cooling in unprecedented detail in a variety of emerging materials. It is, however, not a straightforward task to determine the temperature with high accuracy. This is mainly attributable to an \textit{a priori} unknown position of the Fermi level and the fact that the shape of the Fermi edge can be severely perturbed when the state in question is crossing the Fermi energy. Here, we introduce a method that circumvents these difficulties and accurately extracts both the temperature and the position of the Fermi level for a hot carrier distribution by tracking the occupation statistics of the carriers measured in a TR-ARPES experiment. 
\end{abstract}

\maketitle

\section{I. Introduction}

Angle-resolved photoemission spectroscopy (ARPES) is a well-established and powerful experimental technique that provides high-resolution energy- and momentum-resolved information about the occupied electronic band structure of solid-state materials in thermal equilibrium. In recent years, this techniques has been successfully carried into the time domain by applying various pump-probe schemes with femtosecond time resolution. Simultaneous acquisition of spectral and dynamic information about the out-of-equilibrium carrier excitation and relaxation processes at selected momenta in the Brillouin zone is thereby made possible, opening unprecedented opportunities for understanding the ultrafast transient changes in the charge ordered states in charge density wave materials \cite{Rohwer:2011,Petersen:2011,Hellmann:2012}, the time-scale for the interaction between electrons and bosonic particles in high-temperature superconductors \cite{Perfetti:2007,smallwood:2012} and the dynamics taking place on the Dirac cone in graphene \cite{johannsen:2013,gierz:2013}.

In a time-resolved ARPES (TR-ARPES) experiment an out-of-equilibrium electron distribution is created by pumping with an infrared pump pulse and then measured via photoemission with an ultraviolet probe pulse after a variable time delay. Electron-electron scattering processes on the femtosecond time scale \cite{Boven:2007,Ashcroft:1976} lead to an ultrafast thermalization of the photoexcited electrons to a Fermi-Dirac distribution with a temperature that can be significantly higher than that of the lattice. This population of hot electrons then subsequently cools by transferring its energy to the lattice through electron-phonon scattering events \cite{Anisimov:1974, Allen:1987b}. A great advantage of TR-ARPES lies in the possibility to directly measure the energy- and momentum-dependent carrier distribution, enabling a model-free determination of the temporal evolution of the electronic temperature. In order to arrive at a correct description of the relaxation dynamics of the hot electrons, it is critically important that such a determination is performed quantitatively and with high accuracy. This is evident in recent studies on topological insulators \cite{Crepaldi:2012,Wang:2012d}, superconductors \cite{Perfetti:2007,Dal-Conte:2012,Avigo:2013,Rettig:2013}, graphite \cite{Kampfrath:2005,Stange:2013} and graphene \cite{johannsen:2013,gierz:2013}, which all discuss various cooling mechanisms involving lattice modes. The studies on graphene, furthermore, address the issue of possible carrier multiplication - a process where the number of excited electron-hole pairs exceeds the initially optically generated electron-hole pairs. An accurate determination of the electronic temperature is of paramount importance to quantify this process \cite{johannsen:2013}.     

There has been little discussion thus far in the ARPES literature on how best to extract the temperature directly from the photoemission signal. This is perhaps not so surprising as in equilibrium ARPES it suffices to determine the temperature of the entire sample using either a thermocouple in thermal contact with the sample or from the readings of an infrared pyrometer. The static position of the Fermi level, on the other hand, is typically determined from a polycrystalline sample showing a virtually constant density of states in close vicinity of the Fermi energy. The study by Kr\"{o}ger \emph{et al.} \cite{Kroger:2001} addresses the issue of extracting information about the temperature from an ARPES measurement on a single crystal sample displaying a band crossing at the Fermi level.
They show that the temperature of the sample can be deduced from the incoherently scattered electrons that make up the background signal. This is an intriguing result, relevant for temperature-dependent ARPES experiments where a rapid and reliable determination of the sample temperature is required. For a TR-ARPES experiment, however, the temperature of the ensemble of hot electrons in distinct states in the Brillouin zone, rather than the sample temperature, is the quantity of interest. This temperature can only be extracted by analyzing the distribution of the non-scattered photoemitted hot electrons. In fact, the most direct measure of the electronic temperature is the width of the Fermi-Dirac distribution. As a result, one may be inclined to believe that extracting the temperature of the hot electrons in states showing an unambiguous Fermi level crossing is a relatively easy task once the energy distribution curve (EDC) at the corresponding wave vector has been measured. As Kr\"{o}ger \emph{et al.} \cite{Kroger:2001}, and several references therein, point out, such an approach will, however, not give quantitatively correct values for the position nor width of the Fermi edge owing to several effects including an energy-dependent density of states, many-body interactions and, most importantly, the combination of the band dispersion and the Fermi-cutoff as such. These effects all contribute to a complex line shape and introduce extrinsic background components in the EDC spectrum, see Refs. \cite{Valla:1999b,Kaminski:2001} for examples of this. Matters are not improved by the additional complication that in a TR-ARPES experiment the time-dependent Fermi level position is unknown \emph{a priori}. Consequently, extracting the electronic temperature from TR-ARPES data in a quantitative and reliable manner for any real sample is an involved task that nevertheless calls for a solution given its importance for TR-ARPES studies utilizing this extracted quantity to discuss new physical phenomena. 

Confronted with these challenges, we endeavor to develop a quantitative analysis method, which would allow us to determine the electronic temperature in an accurate and reliable manner. Our method relies on analyzing the statistical distribution of hot electrons along the electronic dispersion procured from constant-energy slices of the spectral function, so-called momentum distribution curves (MDCs). Simulated ARPES intensity data using known parameters and without strong self-energy effects included are utilized for evaluating the accuracy of our MDC analysis method and for performing a comparison with an approach based on EDCs. The particular set of parameters chosen simulates a Dirac-like dispersion as experimentally observed on topological insulators and graphene. As an additional test, we apply these methods to extract the sample temperature from experimental static ARPES data of quasi-free standing monolayer graphene on silicon carbide acquired at a known temperature. We demonstrate that while the EDC approach fails to retrieve the temperature accurately, our analysis based on MDCs is capable of quantitatively reproducing the electronic occupation function and of extracting the electronic temperature as well as the position of the Fermi level (chemical potential) with very high accuracy, even when the influence of instrumental broadening is taken into account.

\section{II. Methods for extracting the electronic temperature}

In the following, we set up and compare two methods for obtaining the electronic temperature from photoemission data. We refer to these as the "EDC method" and the "MDC method", respectively. In order to demonstrate the performance of these two methods, we apply them to simulated photoemission data, which closely emulate our recent TR-ARPES data for hole-doped graphene \cite{johannsen:2013}. We point out that the conclusions drawn later are not restricted to this particular dispersion, but apply more generally.

\subsection{A. Simulations of the spectral function}

We commence by specifying the expression for the photoemission intensity in an ARPES experiment. Within the sudden approximation  \cite{Hufner:2003}, and disregarding matrix element effects and the extrinsic background, the ARPES photoemission intensity is proportional to the hole spectral function $\mathcal{A}$ of the sample multiplied by the Fermi-Dirac distribution $f_{FD}(\omega)$. The product is convoluted with the instrumental resolution functions in energy $G(\Delta \omega)$ and in momentum $G(\Delta k)$, where $\Delta \omega$ and $\Delta k$ are the total energy and momentum resolutions. This yields
\begin{equation} 
\mathcal{I}(\omega,\textbf{k}) = [ \mathcal{A}(\omega,\textbf{k})  f_{FD}(\omega) ] \otimes G(\Delta \omega) \otimes G(\Delta k).
\label{eqn:1}
\end{equation}
The spectral function can be interpreted as the probability of finding an electron with a certain energy and momentum at a given temperature. It is determined by the bare band dispersion $\epsilon({\textbf{k}})$ and the quasiparticle self-energy $\Sigma(\omega)$ = $\Sigma'(\omega)$ + $i\Sigma''(\omega)$, which holds information about the many-body effects. The assumption of a local or momentum-independent self-energy is in many cases a reasonable starting point. By adopting this assumption here, $\mathcal{A}$ takes on the following form:
\begin{figure*}
\includegraphics[width=.99 \columnwidth]{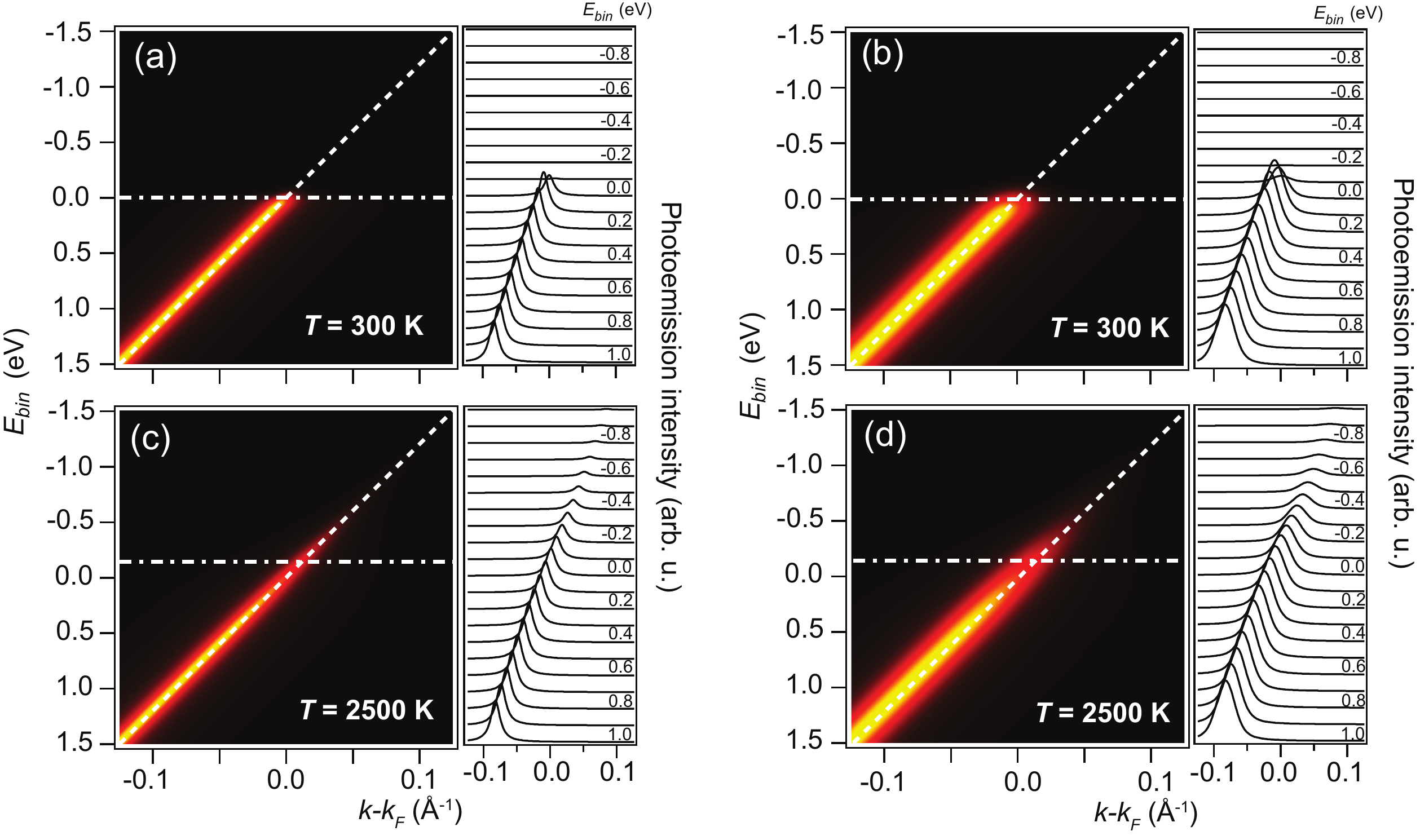}\\
\caption{Simulated spectral function at a temperature of (a)-(b) 300~K and (c)-(d) 2500~K. The spectral functions in (b) and (d) were convoluted with two Gaussians with full-width at half maximum of 0.15~eV and 0.02~\AA$^{-1}$ in energy and momentum, respectively. The simulated ARPES data in (a) and (c), on the other hand, correspond to the non-broadened spectral function. The dashed white line represents the linear bare band and the horisontal dash-dotted line marks the position of the simulated chemical potential. The panels to the right of the spectral function present MDCs for binding energies in the range from 1~eV to -1~eV.}
\label{fig:1}
\end{figure*}
\begin{equation}
 \mathcal{A}(\omega,\textbf{k})=\frac{\pi^{-1}|\Sigma''(\omega)|}{[\hbar \omega-\epsilon({\textbf{k}})-\Sigma'(\omega)]^{2}+\Sigma''(\omega)^{2}}.
\label{eqn:2}
\end{equation}
From Eq. (\ref{eqn:2}), it follows that under the given assumptions a MDC cut through the spectral function should have a Lorentzian line shape plotted against $\textbf{k}-\textbf{k}_F$. Notice also that by neglecting any temperature dependence of the spectral function, the temperature in Eq. (\ref{eqn:1}) enters solely through the Fermi-Dirac distribution $f_{FD}(\omega) = (e^{(\omega - \mu(T))/k_bT}+1)^{-1}$. Disregarding any effects of photodoping and lattice distortions, the temperature-dependent shift of the chemical potential $\mu(T)$ originates from the charge neutrality condition requiring that the total density of electrons is independent of temperature. Since the electron density can be expressed as an integral over the density of states (DOS) multiplied by the Fermi-Dirac distribution \cite{Ashcroft:1976} the behavior of $\mu(T)$ is given once the DOS is known. 

In order to simulate the spectral function, we assume that the bare band $\epsilon({\textbf{k}})$ is a linear function of momentum, as is the case for the low-energy charge carriers in graphene. By setting $\Sigma'(\omega) = 0$, we neglect any renormalization of the bare band. Furthermore, we use a constant broadening $\Sigma''(\omega) = 0.1$~eV, thereby modeling the physical situation where the lifetime of the quasiparticle is limited by impurity scattering only. To simulate the temperature dependence of the chemical potential in Eq. (1), we exploit that the charge carrier concentration is constant for a linear dispersion, and use a DOS linearly increasing with binding energy. This choice corresponds to the situation in hole-doped graphene as described in the supplementary material of Ref. \cite{johannsen:2013}. Note that in this case the chemical potential shifts to lower binding energies as the temperature is raised. To investigate the influence of finite energy and momentum resolution, we consider the case of TR-ARPES experiments based on high harmonic generation (HHG) techniques, where values are typically on the order of 0.15~eV and 0.02~\AA$^{-1}$, respectively, for a time resolution of $\sim$ 50 fs. These values are much larger than the limits of current electron analyzers, which can be reached in static ARPES or low-energy TR-ARPES, because the bandwidth of the high harmonics limits the energy resolution, and the often low statistics in HHG TR-ARPES measurements requires using analyzer slits and lens modes that allow for more counts on the detector at the expense of energy and momentum resolution. The finite instrumental resolution enters in the photoemission intensity in Eq. (\ref{eqn:1}) through a convolution with two Gaussian functions $G(\Delta \omega)$ and $G(\Delta k)$ with unit area and full-width at half maximum (FWHM) of $\Delta \omega$ = 0.15~eV and $\Delta k$~=~0.02~\AA$^{-1}$, respectively. For practical purposes, we assume a constant momentum resolution. 

Simulated spectral functions with and without broadening at temperatures of 300~K and 2500~K are presented in Fig. 1, where the binding energy is referenced to the position of the Fermi level at 300~K. Note that the non-broadened spectral function corresponds to the limit of infinitely high energy- and momentum-resolution. We used 2500~K as a realistic value for the temperature of a hot thermalized population of photoexcited electrons in a TR-ARPES experiment \cite{johannsen:2013,gierz:2013}. The bare dispersion is shown as a white dashed line and is assumed to be measured relative to the Fermi wave vector $\textbf{k}_F$. Fig. 1 also displays a stack of MDCs for each spectral function. Owing to a smeared out Fermi-Dirac distribution of the hot electrons at 2500~K, one clearly observes peaks in the MDC cuts even at small (negative) binding energies in Fig. 1(c)-(d). 

\subsection{B. One-dimensional cuts of the spectral function}

To proceed with the analysis, we start out by employing the EDC method to the simulated ARPES data. With this method, we analyze one-dimensional EDCs of the spectral function binned over a small $k$-window ($\pm 0.01$~\AA$^{-1}$) around $k_F$ simulated for four different temperatures between 300~K and 2500~K, and for non-broadened and broadened data as shown in Fig. 2(a)-(b). All the EDCs display a peak with a tail that extends to lower binding energies. We strive to extract information about the hot carrier statistics from this tail. One immediately notices, however, by comparing the broadened and non-broadened EDCs, that the extension of the tail in the Gaussian broadened data is larger. This effect of the resolution function on the tail of the spectrum indicates that extracting a temperature and chemical potential from the tail of an EDC
in an accurate and reliable manner is a difficult task.

\begin{figure}
\includegraphics[width=.5 \columnwidth]{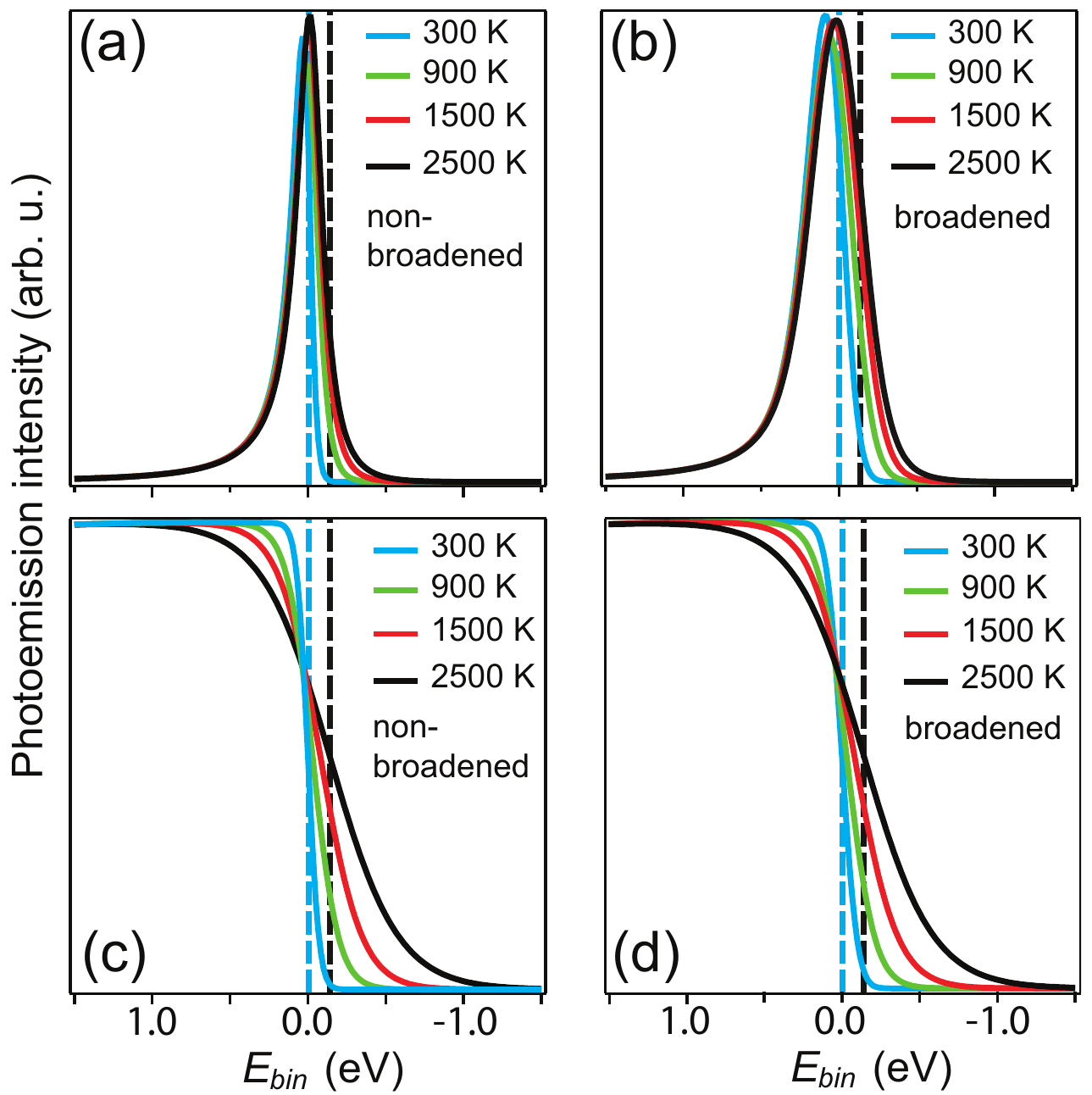}\\
\caption{(a)-(b) EDCs at the Fermi wave vector $k_F$ (binned over a $k$-window of $\pm 0.01$~\AA$^{-1}$) for four different temperatures for (a) non-broadened, and (b) broadened data. (c)-(d) Integral of Lorentzian fits to a stack of MDCs for (c) non-broadened, and (d) broadened case. The vertical dashed lines mark the position of the chemical potential at 300 K (blue) and 2500 K (black).}
\label{fig:2}
\end{figure}

In the MDC method, we slice the spectral function at each binding energy into an MDC, resulting in more than 300 MDCs per image. We then fit the peak shape for each MDC to a Lorentzian function and momentum-integrate the intensity under the resulting Lorentzian line shape. Plotting the value of this integral against binding energy results in the curves shown in Fig. 2(c)-(d) for non-broadened and broadened data. The curves represent the statistical distribution of electrons along the dispersing band. In other words, by tracking the dispersion of the state over the entire binding energy range and integrating the momentum distribution of this state at each energy, we gain access to the Fermi-Dirac distribution of the hot electrons. We observe that the finite resolution does not seem to significantly influence neither the form of the curves nor their extension above the Fermi level. The simulated shift of the Fermi level to lower binding energies is clearly observed as the temperature increases, reaching approximately -150~meV at 2500~K. This behavior reflects the requirement of charge neutrality for our choice of a linearly increasing DOS. Furthermore, because the MDC method tracks the peak along the dispersing band while the EDC method only captures the peak intensity fixed at $k_F$, the tail of the curves in Fig. 2(c)-(d) extends much further than that of the EDCs in Fig. 2(a)-(b) at elevated temperatures. 

\subsection{C. Determining the electronic temperature}

In order to extract the electronic temperature, we fit a Fermi-Dirac distribution to the curves in Fig. 2. For the broadened data, the Fermi-Dirac distribution is convoluted in the fit with a Gaussian line shape having FWHM~=~0.15~eV.  

The fit to the EDCs is performed over a binding energy range from -0.1~eV to -1.5~eV. By doing so, only the tail of the EDC is included in the fit as shown in Fig. 3(a). As expected, we find that the fitted temperature value depends strongly on the range of binding energies chosen in the fit, varying almost 200~K as judged by comparing all reasonably converged fits performed in different binding energy ranges. When the Gaussian broadened Fermi-Dirac function is fitted to the Fermi edge procured from Lorentzians fitted to the MDCs, we observe no significant dependence of the chosen binding energy window on the extracted temperature. In fact, it is possible to fit the entire range from 1.5~eV to -1.5~eV as in Fig. 3(b), where an excellent fit to the data is obtained. An EDC integrated over the entire $k$-range of the simulated data is displayed in Fig. 3(c) along with its Fermi-Dirac fit. In this case all of the simulated intensity is summed instead of just binning the EDC around $k_F$ as in the EDC method. This provides a similarly good fit of the Fermi-Dirac function as the MDC method.

\begin{figure}[t]
\includegraphics[width=.6 \columnwidth]{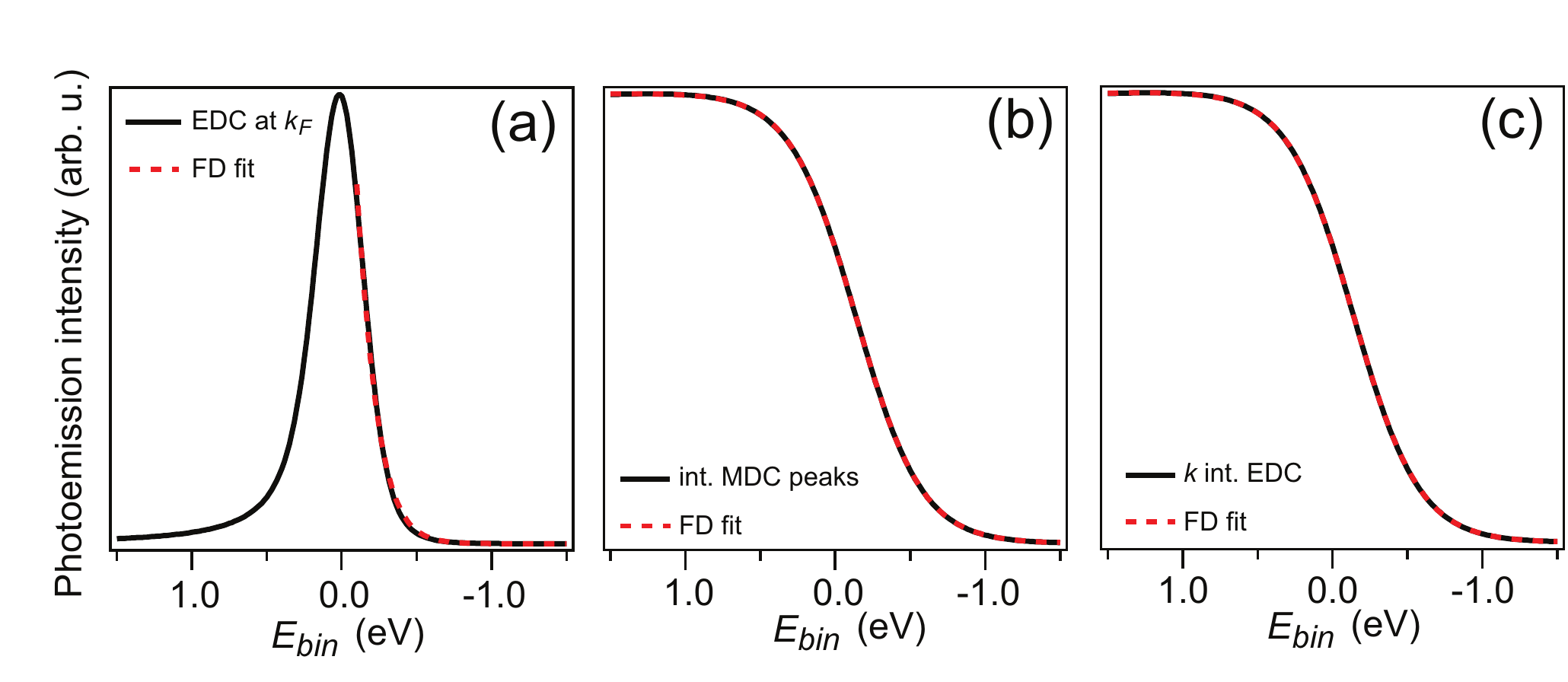}\\
\caption{(a) EDC binned around $k_F$, (b) integral of fitted MDC peak intensities, and (c) $k$-integrated EDC at a temperature of 2500~K. The curves were obtained from the simulated TR-ARPES data displayed in Fig. 1(d). In all cases, the fit was performed with a Fermi-Dirac (FD) distribution convoluted with a Gaussian having FWHM = 0.15~eV.}
\label{fig:3}
\end{figure}

From the fit, we extract the electronic temperature and the value for the chemical potential. Figure 4(a) presents the fitted temperatures plotted against the (real) temperature used to simulate the TR-ARPES data. Interestingly, the MDC method provides a close to perfect one-to-one correspondence between the real and the extracted temperature regardless of the energy- and momentum-broadening of the simulated data. Contrarily, the EDC method substantially underestimates the temperature for all simulated temperatures and does so by more than 60~$\%$ at 2500~K. The extracted values for the chemical potential, shown in Fig. 4(b), are in excellent quantitative agreement with the real values in the MDC method as expected from the excellent fit to the Fermi edge in Fig. 3(b). The EDC method, on the other hand, is not capable of even capturing the qualitative trends in the data. Most severe is the case of the non-broadened data, where the chemical potential even shifts in the wrong direction.

\begin{figure}
\includegraphics[width=0.4 \columnwidth]{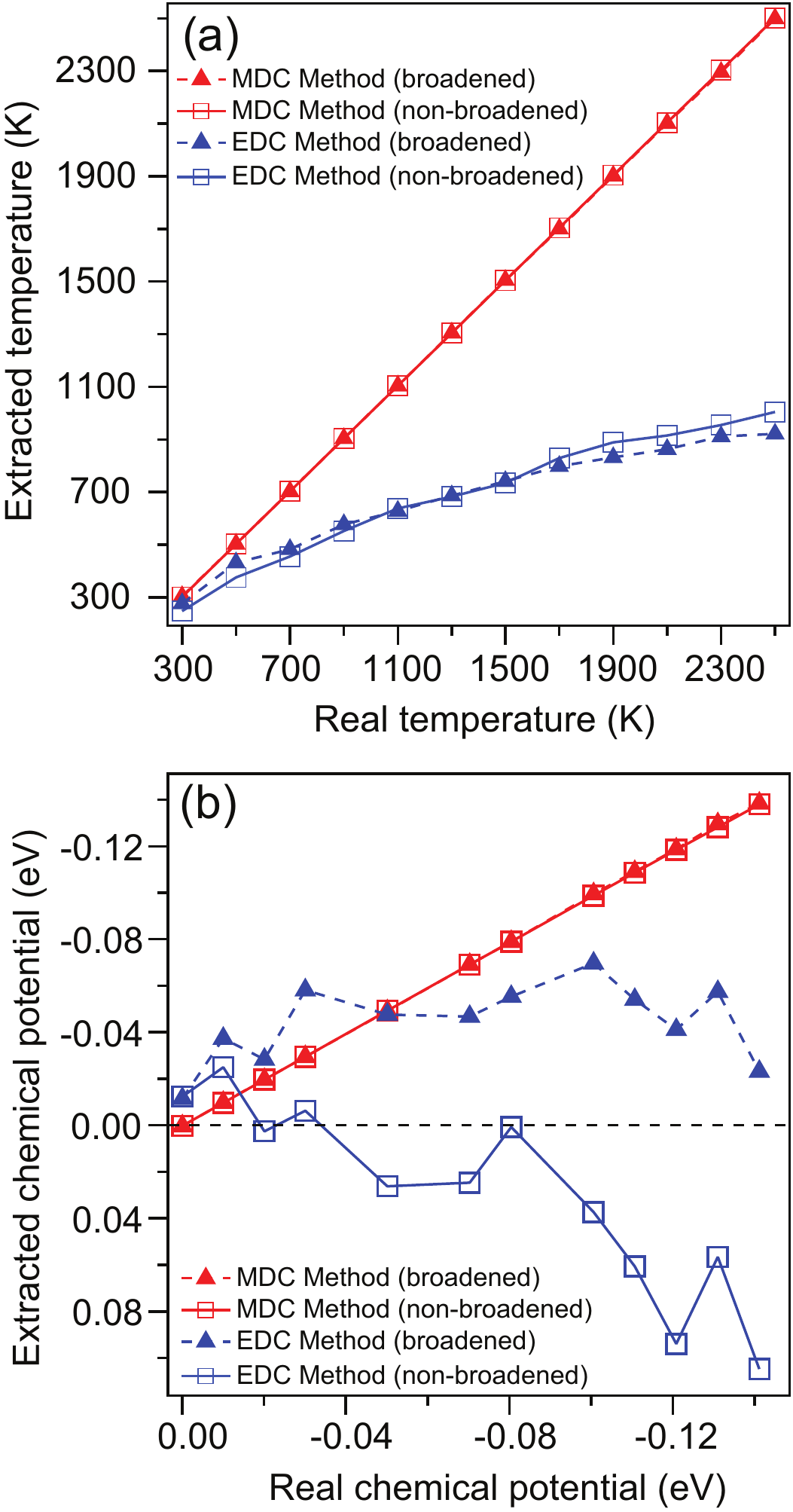}\\
\caption{Plot of fitted versus simulated values for (a) the electronic temperature and (b) the chemical potential. Results are based on both non-broadened and broadened spectral functions.}
\label{fig:4}
\end{figure}

\begin{figure*} 
\includegraphics[width=0.8 \columnwidth]{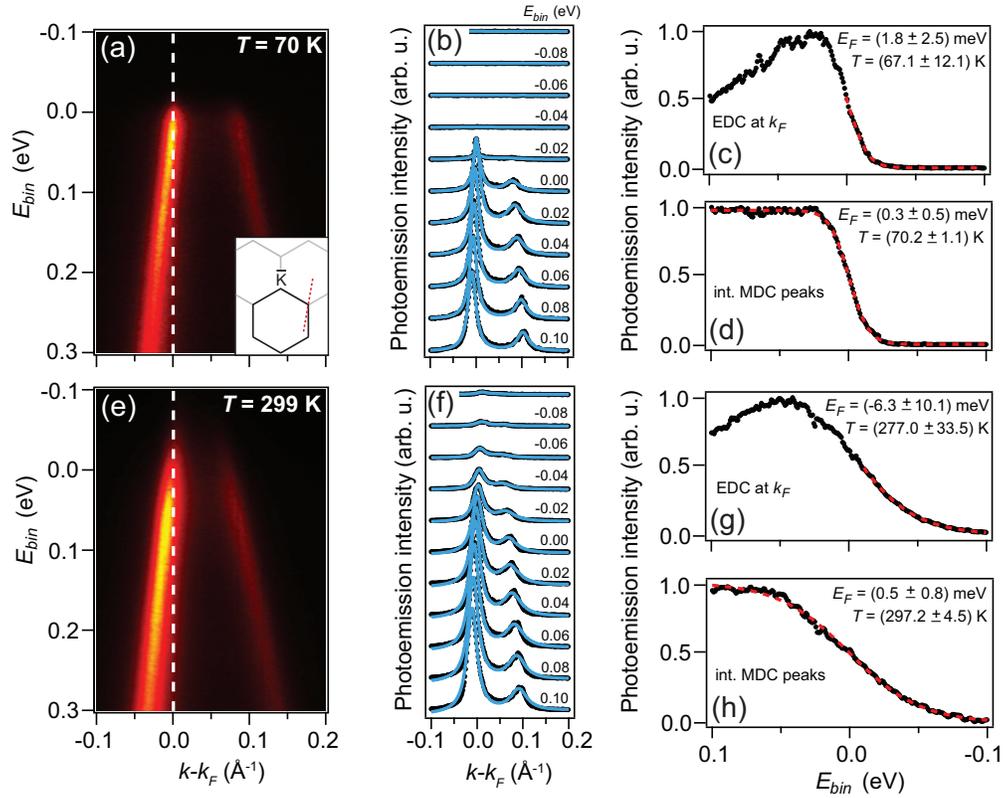}\\
\caption{ARPES measurements of the spectral function of QFMLG on SiC at a sample temperature of (a) 70~K and (e) 299~K along the direction in the Brillouin zone given by the dashed red line in the inset in (a). The data in (a)-(d) correspond to measurements taken at 70~K, and in (e)-(h) to those taken at 299~K. A subset of MDCs in close vicinity of the Fermi level is shown in (b) and (f). Black markers correspond to measured intensity, and blue curves are results of double Lorentzian fits to the data. EDCs at k$_F$ (binned over a $k$-window of $\pm 0.01$~\AA$^{-1}$), marked by the dashed white lines in (a) and (e), are given in (c) and (g). The integrated Lorentzian peak for the most intense branch is plotted as a function of binding energy in (d) and (h). Data points are given as black markers, and Fermi Dirac function fits are shown by red dashed lines. The reported Fermi level and temperature values are results of the fit.}
\label{fig:5}
\end{figure*}

\section{III. Analysis of Experimental ARPES Data}

We now test the performance of the EDC and MDC methods by applying them to experimental static ARPES data obtained from a sample held at a fixed, known temperature. In line with the simulated data, we choose to consider experimental ARPES data from a hole-doped monolayer graphene sheet. More precisely, the measured sample consists of an epitaxial layer of graphene on SiC. The danglings bonds of the SiC surface have been passivated by intercalation of hydrogen giving rise to a quasi free-standing monolayer graphene (QFMLG). The data were acquired at the SGM3 beamline of the synchrotron radiation source ASTRID \cite{Hoffmann:2004} using a photon energy of 32~eV, and with total energy- and $k$-resolution of 13~meV and 0.01~\AA$^{-1}$. The results are published in Ref. \cite{Johannsen:2013} where also further details on the sample and ARPES experiment are given.  We consider two datasets obtained with the sample temperature at 70~K (using a closed cycle liquid helium cryostat) and at 299~K as shown in Fig. 5(a) and 5(e), respectively. The temperatures were measured with a K-type thermocouple in contact with the sample. For QFMLG on SiC it is not possible to study significantly higher temperatures as this would remove hydrogen from below the graphene damaging the sample \cite{Riedl:2009}. The Fermi level position in Fig. 5(a) and 5(e) was determined by fitting the Fermi edge measured on the polycrystalline Ta sample holder. We observe both branches of the graphene Dirac cone, but the left branch is more intense for this particular cut through the cone. For this reason, the Fermi wave vector is referenced to the left branch, and we proceed to extract the temperature of the electrons in this band. 

Following the EDC method described above, we consider an EDC at $k_F$ (binned over a $k$-window of $\pm 0.01$~\AA$^{-1}$) as marked by a dashed white line in Fig. 5(a) and 5(e), and fit the tail of the peak from binding energies of -0.01 eV to -0.1 eV with a Fermi-Dirac function as shown in Fig. 5(c) and 5(g). Using instead the MDC method, we slice each image into MDCs and fit the two branches with Lorentzian peaks as demonstrated in Fig. 5(b) and 5(f). The fitted Lorentzian peak for the left branch is integrated in each MDC fit, and the value of this integral is plotted as a function of binding energy and fitted from 0.1 eV to -0.1 eV with a Fermi-Dirac distribution as seen in Fig. 5(d) and 5(h). In all fits, the Fermi-Dirac distribution is convoluted with a Gaussian having a FWHM of 13~meV to account for the total energy resolution of the experiment. As was the case for the simulated data, the MDC method gives very accurate values for the Fermi level and temperature. The EDC method, on the other hand, yields large error bars in the fits, underestimates the temperature and provides an incorrect Fermi level position. These observations corroborate the results of our simulations in Fig. 4.   

\section{IV. Discussion}

In previous TR-ARPES works on materials with bands crossing the Fermi level, information about the electronic temperature and chemical potential was obtained from Fermi-Dirac (or exponential) distribution function fits to EDCs taken at the Fermi wave vector or binned over a few wave vector values \cite{Perfetti:2007,Crepaldi:2012}. Such an approach, however, is prone to give inaccurate values in the fit. Above, we applied such an EDC method to simulated TR-ARPES data based on a phenomenological description of the spectral function. Our analysis shows that the quality of the fit as well as the values of the extracted quantities strongly depend on the range of binding energies included in the fit. Furthermore, the fitted electronic temperatures and chemical potential deviate extensively from the values used for generating the data. The problem is particularly pronounced at higher temperatures where the hot electron photoemission signal is not properly captured by an EDC at $k_F$. Comparing the temperature extracted from the EDC analysis of the experimental ARPES data with the measured temperature already reveals discrepancies present at low temperatures in good agreement with the results of the simulated data. These observations confirm the concerns stated earlier regarding performing an EDC analysis on states crossing the Fermi level with the aim of extracting values for the temperature and the chemical potential. One could think of modifying the EDC approach to resolve these difficulties by fitting the entire peak with an appropriate analytical function. A major problem with such an approach, however, is the unavailability of a theoretical description of the complex line shape of an EDC that captures the background external to the spectral function and in particular the energy dependence of the self-energy function in a quantitative way. We have attempted to fit the entire line shape displayed in Fig. 3(a) by a fitting function constructed from a convolution between a Gaussian energy-resolution function and the result of a Fermi-Dirac distribution multiplied by a Lorentzian. Irrespective of whether the chosen form for the self-energy was allowed to carry any energy dependence or not, we were not able to achieve a satisfactory fit to the EDC data using this fitting function. In fact, such a fitting function turned out to be very un-stable in practice.
An alternative application of the EDC method involves analyzing EDCs binned over an appreciable range of momenta. We found, as shown in Fig. 3(c), that in the extreme case of integrating the EDCs over the full momentum range, the fitted Fermi-Dirac function and thus the extracted temperature displays a similar degree of agreement with the real temperature as is the case for our MDC method. We note, however, that such an approach integrates out all momentum information in the spectral function and would thus not be applicable for materials having more than one single band at the Fermi level.

The MDC method presented above has clear advantages: Firstly, the assumption of a Lorentzian line shape for the peak in an MDC is often a very good approximation. This is owed to the generally weak dependence of the self-energy on momentum, and the fact that the actual MDC line shape in experimental ARPES data has a simple background and is symmetric up to relatively large binding energies. An intriguing consequence hereof is the possibility to extract the lifetime of the electronic excitations, which is proportional to the FWHM of the fitted Lorentzian peak, simultaneously with the temperature and the chemical potential. Considering the rather poor energy- and momentum-resolution in current TR-ARPES experiments, one should, however, be cautious about ascribing physical significance to these quantities unless one can compensate for the resolution effects in a reliable manner. Interestingly, in a recent work by Levy \emph{et al.} \cite{GLevy:2013} a method is proposed to efficiently account for this. A second advantage is that by performing Lorentzian fits in the MDC method it is possible to single out the contribution to the carrier dynamics in distinct states in the Brillouin zone. This method is thus applicable to systems exhibiting multiple bands crossing the Fermi level. Finally, the most important finding to emerge from this study is the capability of the MDC method of extracting the values for the electronic temperature and chemical potential in a fully quantitative manner. We assessed the effect of finite energy and momentum resolution on the accuracy of this determination, and found no significant deviation from the expected values due to this effect. Similarly, our analysis showed that the MDC method is very reliable even for very steep bands, i.e. bands where the Fermi velocity of the electrons is high, as in graphene and topological insulators.

\section{V. Conclusion}

The purpose of this study was to develop a method to accurately extract the chemical potential and electronic temperature of the hot electrons belonging to distinct states in the Brillouin zone. Two different methods were applied: One based on energy distribution curves (EDCs) and the other on momentum distribution curves (MDCs). We applied these methods to simulated ARPES intensity data based on a phenomenological description of the spectral function. The method based on fitting EDCs at the Fermi wave vector was found incapable of capturing the electronic temperature with satisfactory accuracy, underestimating its value by more than 1000~K at an actual temperature of 2500~K, and provided unphysical behavior of the chemical potential. These issues were not present when using the MDC method. Here, the statistical distribution of the electrons along the dispersing band was obtained by integrating the Lorentzian peak fitted to each MDC cut of the spectral function. By fitting the value of this integral as a function of binding energy to a Fermi-Dirac function, we were able to determine the electronic temperature and chemical potential with high accuracy, even in case of a finite experimental resolution. These findings were confirmed by comparing the measured sample temperature of quasi free-standing graphene on SiC to the temperature extracted by analyzing experimental static ARPES data using both methods. We believe that the suggested MDC method will be useful in the analysis of experimental data obtained with TR-ARPES and hope and trust that it will serve as a basis for future studies using this promising technique.

\section{Acknowledgements}
We acknowledge financial support from the Danish Council for Independent
Research, Technology and Production Sciences, and the VILLUM foundation. Work at Lausanne is funded by the Swiss NSF.
We are thankful to Felix Fromm, Christian Raidel and Thomas Seyller from the Institute of Physics at the Technical University of Chemnitz for providing the hydrogen intercalated monolayer graphene on SiC sample.


\begin{thebibliography}{27}
\expandafter\ifx\csname natexlab\endcsname\relax\def\natexlab#1{#1}\fi
\expandafter\ifx\csname bibnamefont\endcsname\relax
  \def\bibnamefont#1{#1}\fi
\expandafter\ifx\csname bibfnamefont\endcsname\relax
  \def\bibfnamefont#1{#1}\fi
\expandafter\ifx\csname citenamefont\endcsname\relax
  \def\citenamefont#1{#1}\fi
\expandafter\ifx\csname url\endcsname\relax
  \def\url#1{\texttt{#1}}\fi
\expandafter\ifx\csname urlprefix\endcsname\relax\def\urlprefix{URL }\fi
\providecommand{\bibinfo}[2]{#2}
\providecommand{\eprint}[2][]{\url{#2}}


\bibitem{Rohwer:2011}
\bibinfo{author}{\bibfnamefont{T.}~\bibnamefont{Rohwer}},
\bibinfo{author}{\bibfnamefont{S.}~\bibnamefont{Hellmann}},
\bibinfo{author}{\bibfnamefont{M.}~\bibnamefont{Wiesenmayer}},
\bibinfo{author}{\bibfnamefont{C.}~\bibnamefont{Sohrt}},
\bibinfo{author}{\bibfnamefont{A.}~\bibnamefont{Stange}},
\bibinfo{author}{\bibfnamefont{B.}~\bibnamefont{Slomski}},
\bibinfo{author}{\bibfnamefont{A.}~\bibnamefont{Carr}},
\bibinfo{author}{\bibfnamefont{Y.}~\bibnamefont{Liu}},
\bibinfo{author}{\bibfnamefont{L.~M.}~\bibnamefont{Avila}},
\bibinfo{author}{\bibfnamefont{M.}~\bibnamefont{Kall{\"a}ne}},
\bibinfo{author}{\bibfnamefont{S.}~\bibnamefont{Mathias}},
\bibinfo{author}{\bibfnamefont{L.}~\bibnamefont{Kipp}},
\bibinfo{author}{\bibfnamefont{K.}~\bibnamefont{Rossnagel}},
\bibinfo{author}{\bibfnamefont{M.}~\bibnamefont{Bauer}},
  \bibinfo{journal}{Nature} \textbf{\bibinfo{volume}{471}},
  \bibinfo{eid}{490} (\bibinfo{year}{2011}).
  
  
\bibitem{Petersen:2011}
\bibinfo{author}{\bibfnamefont{J.~C.} \bibnamefont{Petersen}},
  \bibinfo{author}{\bibfnamefont{S.}~\bibnamefont{Kaiser}},
  \bibinfo{author}{\bibfnamefont{N.}~\bibnamefont{Dean}},
  \bibinfo{author}{\bibfnamefont{A.}~\bibnamefont{Simoncig}},
  \bibinfo{author}{\bibfnamefont{H.~Y.} \bibnamefont{Liu}},
  \bibinfo{author}{\bibfnamefont{A.~L.} \bibnamefont{Cavalieri}},
  \bibinfo{author}{\bibfnamefont{C.}~\bibnamefont{Cacho}},
  \bibinfo{author}{\bibfnamefont{I.~C.~E.} \bibnamefont{Turcu}},
  \bibinfo{author}{\bibfnamefont{E.}~\bibnamefont{Springate}},
  \bibinfo{author}{\bibfnamefont{F.}~\bibnamefont{Frassetto}},
  \bibnamefont{et~al.}, \bibinfo{journal}{Phys. Rev. Lett.}
  \textbf{\bibinfo{volume}{107}}, \bibinfo{pages}{177402}
  (\bibinfo{year}{2011}).
  
\bibitem{Hellmann:2012}
\bibinfo{author}{\bibfnamefont{S.}~\bibnamefont{Hellmann}},
\bibinfo{author}{\bibfnamefont{T.}~\bibnamefont{Rohwer}},
\bibinfo{author}{\bibfnamefont{M.}~\bibnamefont{Kall{\"a}ne}},
\bibinfo{author}{\bibfnamefont{K.}~\bibnamefont{Hanff}},
\bibinfo{author}{\bibfnamefont{C.}~\bibnamefont{Sohrt}},
\bibinfo{author}{\bibfnamefont{A.}~\bibnamefont{Stange}},
\bibinfo{author}{\bibfnamefont{A.}~\bibnamefont{Carr}},
\bibinfo{author}{\bibfnamefont{M.~M.}~\bibnamefont{Murnane}},
\bibinfo{author}{\bibfnamefont{H.~C.}~\bibnamefont{Kapteyn}},
\bibinfo{author}{\bibfnamefont{L.}~\bibnamefont{Kipp}},
\bibinfo{author}{\bibfnamefont{M.}~\bibnamefont{Bauer}},
\bibinfo{author}{\bibfnamefont{K.}~\bibnamefont{Rossnagel}},
  \bibinfo{journal}{Nature Communications} \textbf{\bibinfo{volume}{3}},
  \bibinfo{eid}{1069} (\bibinfo{year}{2012}).  
  
 \bibitem[{\citenamefont{Perfetti et~al.}(2007)\citenamefont{Perfetti, Loukakos,
  Lisowski, Bovensiepen, Eisaki, and Wolf}}]{Perfetti:2007}
\bibinfo{author}{\bibfnamefont{L.}~\bibnamefont{Perfetti}},
  \bibinfo{author}{\bibfnamefont{P.~A.} \bibnamefont{Loukakos}},
  \bibinfo{author}{\bibfnamefont{M.}~\bibnamefont{Lisowski}},
  \bibinfo{author}{\bibfnamefont{U.}~\bibnamefont{Bovensiepen}},
  \bibinfo{author}{\bibfnamefont{H.}~\bibnamefont{Eisaki}}, \bibnamefont{and}
  \bibinfo{author}{\bibfnamefont{M.}~\bibnamefont{Wolf}},
  \bibinfo{journal}{Phys. Rev. Lett.} \textbf{\bibinfo{volume}{99}},
  \bibinfo{pages}{197001} (\bibinfo{year}{2007}).
  
 \bibitem{smallwood:2012}
\bibinfo{author}{\bibfnamefont{C.~L.}~\bibnamefont{Smallwood}},
\bibinfo{author}{\bibfnamefont{J.~P.}~\bibnamefont{Hinton}},
\bibinfo{author}{\bibfnamefont{C.}~\bibnamefont{Jozwiak}},
\bibinfo{author}{\bibfnamefont{W.}~\bibnamefont{Zhang}},
\bibinfo{author}{\bibfnamefont{J.~D.}~\bibnamefont{Koralek}},
\bibinfo{author}{\bibfnamefont{H.}~\bibnamefont{Eisaki}},
\bibinfo{author}{\bibfnamefont{D.-H.}~\bibnamefont{Lee}},
\bibinfo{author}{\bibfnamefont{J.}~\bibnamefont{Orenstein}},
\bibinfo{author}{\bibfnamefont{A.}~\bibnamefont{Lanzara}},
  \bibinfo{journal}{Science} \textbf{\bibinfo{volume}{336}},
  \bibinfo{eid}{1137} (\bibinfo{year}{2012}).
  
\bibitem{johannsen:2013}
\bibinfo{author}{\bibfnamefont{J.~C.} \bibnamefont{Johannsen}},
  \bibinfo{author}{\bibfnamefont{S.}~\bibnamefont{Ulstrup}},
  \bibinfo{author}{\bibfnamefont{F.}~\bibnamefont{Cilento}},
  \bibinfo{author}{\bibfnamefont{A.}~\bibnamefont{Crepaldi}},
  \bibinfo{author}{\bibfnamefont{M.} \bibnamefont{Zacchigna}},
  \bibinfo{author}{\bibfnamefont{C.}~\bibnamefont{Cacho}},
  \bibinfo{author}{\bibfnamefont{I.~C.~E.} \bibnamefont{Turcu}},
  \bibinfo{author}{\bibfnamefont{E.}~\bibnamefont{Springate}},
  \bibinfo{author}{\bibfnamefont{F.}~\bibnamefont{Fromm}},
  \bibinfo{author}{\bibfnamefont{C.}~\bibnamefont{Raidel}},
   \bibinfo{author}{\bibfnamefont{T.}~\bibnamefont{Seyller}},
   \bibinfo{author}{\bibfnamefont{F.}~\bibnamefont{Parmigiani}},
   \bibinfo{author}{\bibfnamefont{M.}~\bibnamefont{Grioni}},
   \bibinfo{author}{\bibfnamefont{P.}~\bibnamefont{Hofmann}},
   \bibinfo{journal}{Phys. Rev. Lett.}
  \textbf{\bibinfo{volume}{111}}, \bibinfo{pages}{027403}
  (\bibinfo{year}{2013}).	

\bibitem{gierz:2013}
  \bibinfo{author}{\bibfnamefont{I.}~\bibnamefont{Gierz}},
\bibinfo{author}{\bibfnamefont{J.~C.} \bibnamefont{Petersen}},
  \bibinfo{author}{\bibfnamefont{M.}~\bibnamefont{Mitrano}},
    \bibinfo{author}{\bibfnamefont{C.}~\bibnamefont{Cacho}},
  \bibinfo{author}{\bibfnamefont{I.~C.~E.} \bibnamefont{Turcu}},
  \bibinfo{author}{\bibfnamefont{E.}~\bibnamefont{Springate}},
  \bibinfo{author}{\bibfnamefont{A.}~\bibnamefont{St\"{o}hr}},
  \bibinfo{author}{\bibfnamefont{A.}~\bibnamefont{K\"{o}hler}},
  \bibinfo{author}{\bibfnamefont{U.} \bibnamefont{Starke}},
  \bibinfo{author}{\bibfnamefont{A.} \bibnamefont{Cavalleri}},
 \bibinfo{journal}{Nat. Mater.}
  (\bibinfo{year}{2013}).
  

\bibitem{Boven:2007}
\bibinfo{author}{\bibfnamefont{U.}~\bibnamefont{Bovensiepen}},
  \bibinfo{journal}{Journal of Physics: Condensed Matter} \textbf{\bibinfo{volume}{19}},
  \bibinfo{eid}{083201} (\bibinfo{year}{2007}).  


\bibitem{Ashcroft:1976}
\bibinfo{author}{\bibfnamefont{N.~E.}~\bibnamefont{Ashcroft}},
\bibinfo{author}{\bibfnamefont{N.~D.}~\bibnamefont{Mermin}},
  \bibinfo{book}{Solid state physics},
  \bibinfo{publisher}{Saunders College},
  \bibinfo{Address}{Philadelphia},
  \bibinfo{edition}{International ed.} (\bibinfo{year}{1976}),

\bibitem{Anisimov:1974}
\bibinfo{author}{\bibfnamefont{S.~I.}~\bibnamefont{Anisimov}},
\bibinfo{author}{\bibfnamefont{B.~L.}~\bibnamefont{Kapeliovich}},
\bibinfo{author}{\bibfnamefont{T.~L.}~\bibnamefont{Perel'man}},
  \bibinfo{journal}{J. Exp. Theor. Phys.} \textbf{\bibinfo{volume}{66}},
  \bibinfo{eid}{776} (\bibinfo{year}{1974}).

\bibitem{Allen:1987b}
\bibinfo{author}{\bibfnamefont{P.~B.}~\bibnamefont{Allen}},
  \bibinfo{journal}{Phys. Rev. Lett.} \textbf{\bibinfo{volume}{59}},
  \bibinfo{eid}{1460} (\bibinfo{year}{1987}).

\bibitem{Crepaldi:2012}
  \bibinfo{author}{\bibfnamefont{A.}~\bibnamefont{Crepaldi}},
   \bibinfo{author}{\bibfnamefont{B.}~\bibnamefont{Ressel}},
   \bibinfo{author}{\bibfnamefont{F.}~\bibnamefont{Cilento}},
  \bibinfo{author}{\bibfnamefont{M.} \bibnamefont{Zacchigna}},
  \bibinfo{author}{\bibfnamefont{C.} \bibnamefont{Grazioli}},
  \bibinfo{author}{\bibfnamefont{H.}~\bibnamefont{Berger}},
  \bibinfo{author}{\bibfnamefont{P.}~\bibnamefont{Bugnon}},
  \bibinfo{author}{\bibfnamefont{K.}~\bibnamefont{Kern}},
  \bibinfo{author}{\bibfnamefont{M.}~\bibnamefont{Grioni}},
  \bibinfo{author}{\bibfnamefont{F.}~\bibnamefont{Parmigiani}},
   \bibinfo{journal}{Phys. Rev. B}
  \textbf{\bibinfo{volume}{86}}, \bibinfo{pages}{205133}
  (\bibinfo{year}{2012}).
  
\bibitem{Wang:2012d}
  \bibinfo{author}{\bibfnamefont{Y.~H.}~\bibnamefont{Wang}},
\bibinfo{author}{\bibfnamefont{D.} \bibnamefont{Hsieh}},
  \bibinfo{author}{\bibfnamefont{E.~J.}~\bibnamefont{Sie}},
  \bibinfo{author}{\bibfnamefont{H.}~\bibnamefont{Steinberg}},
  \bibinfo{author}{\bibfnamefont{D.~R.} \bibnamefont{Gardner}},
  \bibinfo{author}{\bibfnamefont{Y.~S.}~\bibnamefont{Lee}},
  \bibinfo{author}{\bibfnamefont{P.}~\bibnamefont{Jarillo-Herrero}},
  \bibinfo{author}{\bibfnamefont{N.}~\bibnamefont{Gedik}},
   \bibinfo{journal}{Phys. Rev. Lett.}
  \textbf{\bibinfo{volume}{109}}, \bibinfo{pages}{127401}
  (\bibinfo{year}{2012}).
  
\bibitem[{\citenamefont{Dal~Conte et~al.}(2012)\citenamefont{Dal~Conte,
  Giannetti, Coslovich, Cilento, Bossini, Abebaw, Banfi, Ferrini, Eisaki,
  Greven et~al.}}]{Dal-Conte:2012}
\bibinfo{author}{\bibfnamefont{S.}~\bibnamefont{Dal~Conte}},
  \bibinfo{author}{\bibfnamefont{C.}~\bibnamefont{Giannetti}},
  \bibinfo{author}{\bibfnamefont{G.}~\bibnamefont{Coslovich}},
  \bibinfo{author}{\bibfnamefont{F.}~\bibnamefont{Cilento}},
  \bibinfo{author}{\bibfnamefont{D.}~\bibnamefont{Bossini}},
  \bibinfo{author}{\bibfnamefont{T.}~\bibnamefont{Abebaw}},
  \bibinfo{author}{\bibfnamefont{F.}~\bibnamefont{Banfi}},
  \bibinfo{author}{\bibfnamefont{G.}~\bibnamefont{Ferrini}},
  \bibinfo{author}{\bibfnamefont{H.}~\bibnamefont{Eisaki}},
  \bibinfo{author}{\bibfnamefont{M.}~\bibnamefont{Greven}},
  \bibnamefont{et~al.}, \bibinfo{journal}{Science}
  \textbf{\bibinfo{volume}{335}}, \bibinfo{pages}{1600} (\bibinfo{year}{2012}).
  
\bibitem{Avigo:2013}
\bibinfo{author}{\bibfnamefont{I.}~\bibnamefont{Avigo}},
\bibinfo{author}{\bibfnamefont{R.}~\bibnamefont{Cort\'es}},
\bibinfo{author}{\bibfnamefont{L.}~\bibnamefont{Rettig}},
\bibinfo{author}{\bibfnamefont{S.}~\bibnamefont{Thirupathaiah}},
\bibinfo{author}{\bibfnamefont{H.~S.}~\bibnamefont{Jeevan}},
\bibinfo{author}{\bibfnamefont{P.}~\bibnamefont{Gegenwart}},
\bibinfo{author}{\bibfnamefont{T.}~\bibnamefont{Wolf}},
\bibinfo{author}{\bibfnamefont{M.}~\bibnamefont{Ligges}},
\bibinfo{author}{\bibfnamefont{M.}~\bibnamefont{Wolf}},
\bibinfo{author}{\bibfnamefont{J.}~\bibnamefont{Fink}},
\bibinfo{author}{\bibfnamefont{U.}~\bibnamefont{Bovensiepen}},
  \bibinfo{journal}{J. Phys.: Condens. Matter} \textbf{\bibinfo{volume}{25}},
  \bibinfo{eid}{094003} (\bibinfo{year}{2013}).  
	
\bibitem{Rettig:2013}
\bibinfo{author}{\bibfnamefont{L.}~\bibnamefont{Rettig}},
\bibinfo{author}{\bibfnamefont{R.}~\bibnamefont{Cort\'es}},
\bibinfo{author}{\bibfnamefont{H.~S.}~\bibnamefont{Jeevan}},
\bibinfo{author}{\bibfnamefont{P.}~\bibnamefont{Gegenwart}},
\bibinfo{author}{\bibfnamefont{T.}~\bibnamefont{Wolf}},
\bibinfo{author}{\bibfnamefont{J.}~\bibnamefont{Fink}},
\bibinfo{author}{\bibfnamefont{U.}~\bibnamefont{Bovensiepen}},
  \bibinfo{journal}{New J. Phys.} \textbf{\bibinfo{volume}{15}},
  \bibinfo{eid}{083023} (\bibinfo{year}{2013}).  

\bibitem[{\citenamefont{Kampfrath et~al.}(2005)\citenamefont{Kampfrath,
  Perfetti, Schapper, Frischkorn, and Wolf}}]{Kampfrath:2005}
\bibinfo{author}{\bibfnamefont{T.}~\bibnamefont{Kampfrath}},
  \bibinfo{author}{\bibfnamefont{L.}~\bibnamefont{Perfetti}},
  \bibinfo{author}{\bibfnamefont{F.}~\bibnamefont{Schapper}},
  \bibinfo{author}{\bibfnamefont{C.}~\bibnamefont{Frischkorn}},
  \bibnamefont{and} \bibinfo{author}{\bibfnamefont{M.}~\bibnamefont{Wolf}},
  \bibinfo{journal}{Phys. Rev. Lett.} \textbf{\bibinfo{volume}{95}},
  \bibinfo{pages}{187403} (\bibinfo{year}{2005}).

\bibitem{Stange:2013}
\bibinfo{author}{\bibfnamefont{A.}~\bibnamefont{Stange}},
\bibinfo{author}{\bibfnamefont{C.}~\bibnamefont{Sohrt}},
\bibinfo{author}{\bibfnamefont{T.}~\bibnamefont{Rohwer}},
\bibinfo{author}{\bibfnamefont{S.}~\bibnamefont{Hellmann}},
\bibinfo{author}{\bibfnamefont{G.}~\bibnamefont{Rohde}},
\bibinfo{author}{\bibfnamefont{L.}~\bibnamefont{Kipp}},
\bibinfo{author}{\bibfnamefont{K.}~\bibnamefont{Rossnagel}},
\bibinfo{author}{\bibfnamefont{M.}~\bibnamefont{Bauer}},
  \bibinfo{journal}{EPJ Web of Conferences} \textbf{\bibinfo{volume}{41}},
  \bibinfo{eid}{04022} (\bibinfo{year}{2013}).

  
 \bibitem{Kroger:2001}
\bibinfo{author}{\bibfnamefont{J.}~\bibnamefont{Kr{{\"o}}ger}},
  \bibinfo{author}{\bibfnamefont{T.}~\bibnamefont{Greber}},
  \bibinfo{author}{\bibfnamefont{T.~J.}~\bibnamefont{Kreutz}},
  \bibinfo{author}{\bibfnamefont{J.}~\bibnamefont{Osterwalder}},
  \bibinfo{journal}{Journal of Electron Spectroscopy and Related Phenomena}
  \textbf{\bibinfo{volume}{113}}, \bibinfo{pages}{241} (\bibinfo{year}{2001}).

\bibitem{Valla:1999b}
\bibinfo{author}{\bibfnamefont{T.}~\bibnamefont{Valla}},
\bibinfo{author}{\bibfnamefont{A.~V.}~\bibnamefont{Fedorov}},
\bibinfo{author}{\bibfnamefont{P.~D.}~\bibnamefont{Johnson}},
\bibinfo{author}{\bibfnamefont{B.~O.}~\bibnamefont{Wells}},
\bibinfo{author}{\bibfnamefont{S.~L.}~\bibnamefont{Hulbert}},
\bibinfo{author}{\bibfnamefont{Q.}~\bibnamefont{Li}},
\bibinfo{author}{\bibfnamefont{G.~D.}~\bibnamefont{Gu}},
\bibinfo{author}{\bibfnamefont{N.}~\bibnamefont{Koshizuka}},
  \bibinfo{journal}{Science} \textbf{\bibinfo{volume}{285}},
  \bibinfo{eid}{2110} (\bibinfo{year}{1999}).  	


\bibitem{Kaminski:2001}
\bibinfo{author}{\bibfnamefont{A.}~\bibnamefont{Kaminski}},
\bibinfo{author}{\bibfnamefont{M.}~\bibnamefont{Randeria}},
\bibinfo{author}{\bibfnamefont{J.~C.}~\bibnamefont{Campuzano}},
\bibinfo{author}{\bibfnamefont{M.~R.}~\bibnamefont{Norman}},
\bibinfo{author}{\bibfnamefont{H.}~\bibnamefont{Fretwell}},
\bibinfo{author}{\bibfnamefont{J.}~\bibnamefont{Mesot}},
\bibinfo{author}{\bibfnamefont{T.}~\bibnamefont{Sato}},
\bibinfo{author}{\bibfnamefont{T.}~\bibnamefont{Takahashi}},
\bibinfo{author}{\bibfnamefont{K.}~\bibnamefont{Kadowaki}},
  \bibinfo{journal}{Physical Review Letters} \textbf{\bibinfo{volume}{86}},
  \bibinfo{eid}{1070} (\bibinfo{year}{2001}).  
	
\bibitem{Hufner:2003}
\bibinfo{author}{\bibfnamefont{S.}~\bibnamefont{H{{\"u}}fner}},
  \bibinfo{book}{Photoelectron spectroscopy},
  \bibinfo{publisher}{Springer},
  \bibinfo{Address}{Berlin},
  \bibinfo{edition}{3rd ed.} (\bibinfo{year}{2003}),

  
  \bibitem[{\citenamefont{Hoffmann et~al.}(2004)\citenamefont{Hoffmann,
  S{\o}ndergaard, Schultz, Li, and Hofmann}}]{Hoffmann:2004}
\bibinfo{author}{\bibfnamefont{S.~V.} \bibnamefont{Hoffmann}},
  \bibinfo{author}{\bibfnamefont{C.}~\bibnamefont{S{\o}ndergaard}},
  \bibinfo{author}{\bibfnamefont{C.}~\bibnamefont{Schultz}},
  \bibinfo{author}{\bibfnamefont{Z.}~\bibnamefont{Li}}, \bibnamefont{and}
  \bibinfo{author}{\bibfnamefont{P.}~\bibnamefont{Hofmann}},
  \bibinfo{journal}{Nuclear Instruments and Methods in Physics Research, A}
  \textbf{\bibinfo{volume}{523}}, \bibinfo{pages}{441} (\bibinfo{year}{2004}).

\bibitem[{\citenamefont{Johannsen et~al.}(2013)\citenamefont{Johannsen,
  Ulstrup, Bianchi, Hatch, Guan, Mazzola, Hornek{\ae}r, Fromm, Raidel, Seyller
  et~al.}}]{Johannsen:2013}
\bibinfo{author}{\bibfnamefont{J.~C.} \bibnamefont{Johannsen}},
  \bibinfo{author}{\bibfnamefont{S.} \bibnamefont{Ulstrup}},
  \bibinfo{author}{\bibfnamefont{M.}~\bibnamefont{Bianchi}},
  \bibinfo{author}{\bibfnamefont{R.}~\bibnamefont{Hatch}},
  \bibinfo{author}{\bibfnamefont{D.}~\bibnamefont{Guan}},
  \bibinfo{author}{\bibfnamefont{F.}~\bibnamefont{Mazzola}},
  \bibinfo{author}{\bibfnamefont{L.}~\bibnamefont{Hornek{\ae}r}},
  \bibinfo{author}{\bibfnamefont{F.}~\bibnamefont{Fromm}},
  \bibinfo{author}{\bibfnamefont{C.}~\bibnamefont{Raidel}},
  \bibinfo{author}{\bibfnamefont{T.}~\bibnamefont{Seyller}},
  \bibnamefont{et~al.}, \bibinfo{journal}{Journal of Physics: Condensed Matter}
  \textbf{\bibinfo{volume}{25}}, \bibinfo{pages}{094001}
  (\bibinfo{year}{2013}).  

\bibitem[{\citenamefont{Riedl et~al.}(2009)\citenamefont{Riedl, Coletti,
  Iwasaki, Zakharov, and Starke}}]{Riedl:2009}
\bibinfo{author}{\bibfnamefont{C.}~\bibnamefont{Riedl}},
  \bibinfo{author}{\bibfnamefont{C.}~\bibnamefont{Coletti}},
  \bibinfo{author}{\bibfnamefont{T.}~\bibnamefont{Iwasaki}},
  \bibinfo{author}{\bibfnamefont{A.~A.} \bibnamefont{Zakharov}},
  \bibnamefont{and} \bibinfo{author}{\bibfnamefont{U.}~\bibnamefont{Starke}},
  \bibinfo{journal}{Physical Review Letters} \textbf{\bibinfo{volume}{103}},
  \bibinfo{eid}{246804} (pages~\bibinfo{numpages}{4}) (\bibinfo{year}{2009}).

\bibitem{GLevy:2013}
\bibinfo{author}{\bibfnamefont{G.}~\bibnamefont{Levy}},
\bibinfo{author}{\bibfnamefont{W.}~\bibnamefont{Nettke}},
\bibinfo{author}{\bibfnamefont{B.~M.}~\bibnamefont{Ludbrook}},
\bibinfo{author}{\bibfnamefont{C.~N.}~\bibnamefont{Veenstra}},
\bibinfo{author}{\bibfnamefont{A.}~\bibnamefont{Damascelli}},
  \bibinfo{journal}{arXiv:1306.2667}(\bibinfo{year}{2013}).  

\end{thebibliography}

\end{document}